\documentclass[sigconf,nonacm]{acmart} %% FOR PREPRINT

\setcopyright{acmlicensed}
\copyrightyear{2025}
\acmYear{2025}
\acmConference[ICSE '26]{}{April 12--18,
  2026}{Rio de Janeiro, BR}

\usepackage{placeins}

\usepackage{geometry}      % For wide tables in landscape mode
\usepackage{booktabs}      % For professional table lines
\usepackage{multirow}      % For cells that span multiple rows
\usepackage{siunitx}       % For aligning numbers on the decimal point
\usepackage{graphicx}      % To rotate text in column headers
\usepackage{tcolorbox}
\definecolor{greenanswerbox}{RGB}{220, 255, 220} % Light green color
\begin{document}

\title{SWEnergy: An Empirical Study on Energy Efficiency in Agentic Issue Resolution Frameworks with SLMs}

% --- Author 1 ---
\author{Arihant Tripathy}
\affiliation{%
  \institution{SERC, IIIT-Hyderabad}
  \city{Hyderabad}
  \state{Telangana}
  \country{India}
}
\email{arihant.tripathy@research.iiit.ac.in}

% --- Author 2 ---
\author{Ch Pavan Harshit}
\affiliation{%
  \institution{SERC, IIIT-Hyderabad}
  \city{Hyderabad}
  \state{Telangana}
  \country{India}
}
\email{pavan.harshit@research.iiit.ac.in}

% --- Author 3 ---
\author{Karthik Vaidhyanathan}
\affiliation{%
  \institution{SERC, IIIT-Hyderabad}
  \city{Hyderabad}
  \state{Telangana}
  \country{India}
}
\email{karthik.vaidhyanathan@iiit.ac.in}

\begin{abstract}
\noindent\textbf{Context.} Autonomous agents powered by Large Language Models (LLMs) are increasingly used for software engineering, but their reliance on large, proprietary models limits deployment on local hardware. This has spurred interest in Small Language Models (SLMs), but their practical effectiveness and efficiency within complex agentic frameworks for automated issue resolution remain poorly understood.

\noindent\textbf{Goal.} We investigate the performance, energy efficiency, and resource consumption of four leading agentic issue resolution frameworks when deliberately constrained to using SLMs. Our goal is to understand the viability of these systems for this task in resource-limited settings and characterize the resulting trade-offs.

\noindent\textbf{Method.} We conduct a controlled evaluation of four leading agentic frameworks (SWE-Agent, OpenHands, Mini SWE Agent, AutoCodeRover) using two SLMs (Gemma-3 4B, Qwen-3 1.7B) on the SWE-bench Verified Mini benchmark. On fixed hardware, we measure energy, duration, token usage, and memory over 150 runs per configuration.

\noindent\textbf{Results.} We find that framework architecture is the primary driver of energy consumption. The most energy-intensive framework, AutoCodeRover (Gemma), consumed 9.4x more energy on average than the least energy-intensive, OpenHands (Gemma). However, this energy is largely wasted. Task resolution rates were near-zero (4\% for AutoCodeRover, 0\% for all others), demonstrating that current frameworks, when paired with SLMs, consume significant energy on unproductive reasoning loops. The SLM's limited reasoning was the bottleneck for \textit{success}, but the framework's design was the bottleneck for \textit{efficiency}.

\noindent\textbf{Conclusions.} Current agentic frameworks, designed for powerful LLMs, fail to operate efficiently with SLMs. We find that framework architecture is the primary driver of energy consumption, but this energy is largely wasted due to the SLMs' limited reasoning. Achieving viable, low-energy solutions requires a paradigm shift from passive orchestration to new architectures that actively manage the SLM's weaknesses.
\end{abstract}




% \received{7 November 2025}
% \received[accepted]{8 December 2025}
\maketitle 

\section{Introduction}

Autonomous software engineering agents, powered by Large Language Models (LLMs), have emerged as a transformative paradigm in software development \cite{10.1145/3712003, 10.1007/s10515-025-00544-2}, demonstrating impressive capabilities in resolving real-world code issues on benchmarks like SWE-bench. However, this success comes at a cost: massive, cloud-hosted models create significant barriers to local deployment due to their high computational costs and energy consumption \cite{10.1145/3727200.3727217}. As the environmental impact of AI systems has become a central concern in the push toward sustainable software engineering practices \cite{10.1145/3743095.3743099}, resource efficiency, in terms of computation, cost, and power consumption, has emerged as a critical factor for widespread adoption.

The challenge of resource efficiency has intensified interest in Small Language Models (SLMs)—open-weight models with billions, rather than hundreds of billions, of parameters \cite{qwenfunc2025, gemma2024}, offering a path toward democratizing autonomous agents via consumer-grade hardware, albeit with inherent limitations in reasoning and instruction-following capabilities. The effectiveness of these SLMs within complex agentic frameworks remains largely unmeasured. An agentic framework defines how a language model structures its reasoning, coordinates external tools, and executes problem-solving workflows \cite{liu2025agent, 10.1007/978-3-032-04403-7_5}. It remains uncertain whether existing frameworks, originally tuned to leverage the capabilities of large models, can maintain their effectiveness when driven by more limited ones.

Agentic frameworks extend beyond single-turn code generation by enabling language models to autonomously plan, execute, and verify the sequence of steps required to resolve software issues. Systems such as \textit{SWE-Agent}, \textit{OpenHands}, and \textit{AutoCodeRover} have demonstrated strong performance on benchmarks like \textit{SWE-bench}, which evaluates real-world GitHub issues as problem statements~\cite{yang2024sweagent, wang2025openhands, zhang2024autocoderover, jimenez2024swebench}. While typically built around large proprietary models, emerging work suggests SLMs may offer resource-efficient alternatives for narrow, well-defined agentic tasks \cite{belcak2025small, kavathekar2025smallmodelsbigtasks}. However, it remains unclear whether existing agentic architectures, originally optimized for powerful LLMs, can effectively utilize SLMs without significant degradation in performance.

Moreover, the energy consumption of multi-turn agentic workflows remains unmeasured, particularly when constrained to SLMs on local hardware. Which raises some important questions such as how much energy do these systems consume in resource-limited settings? How do different architectures influence energy efficiency and task success under these constraints? What trade-offs emerge when accuracy is no longer guaranteed?

As autonomous agents become prevalent and as interest grows in SLMs as a more sustainable alternative to LLMs,  this study addresses these questions by exploring the viability of SLMs in agentic frameworks through a systematic, hardware-level energy evaluation. We position our work not as a direct benchmark but as an investigation into how these agentic frameworks, combined with SLM capabilities, jointly impact resource consumption under realistic constraints of using SLMs on local hardware. Unlike LLM-driven systems, where resource usage is often opaque and hidden behind API abstractions, our setup enables transparent, reproducible measurement of actual CPU and GPU energy consumption. This reframes the evaluation around the joint consideration of correctness and efficiency in resource-limited settings, a critical step toward greening AI-enabled software engineering practices \cite{10.1145/3743095.3743099}.

The main contributions of this study are: (i) a reproducible methodology for measuring the energy consumption of agentic systems running SLMs, (ii) an empirical characterization of the performance-efficiency frontier for SLM-powered agents, revealing that current frameworks are ill-equipped to handle their limitations, and (iii) actionable guidance for designing future agentic systems that can effectively leverage smaller, more accessible models.

\section{Related Work}
\label{sec:related_work}

Research on the resource footprint of AI systems in software engineering generally follows two complementary directions: (1) quantifying the operational cost of running AI models, and (2) assessing the energy efficiency of the code produced by those models.

The first direction focuses on the energy and carbon costs of large-scale language models, revealing their substantial operational and environmental impact~\cite{luccioni2023bloomfootprint, 10.1145/3727200.3727217}.
These findings have motivated a growing interest in SLMs, which enable lightweight, local deployment and lower hardware requirements.
Recent discussions advocate for a shift toward smaller, specialized models, arguing that deploying massive general-purpose models for narrow agentic tasks is an inefficient use of computational resources~\cite{belcak2025small}.

The second direction examines the efficiency of the \emph{outputs} generated by such models.
Multiple studies report that human-written programs tend to be more energy-efficient than those generated by systems like Code Llama or ChatGPT~\cite{apsan2025generating, 27ff5879d4ed4c26a9bfd753652173fe, 10734439, treude2025generativeaiempiricalsoftware}. However, these works primarily evaluate static outputs and do not account for the energy cost of the problem-solving \emph{process}, i.e., the multi-step reasoning, planning, and repair cycles typical in agentic workflows.

To our knowledge, limited research has explored the energy efficiency of issue resolution when performed by autonomous agentic systems.
This gap is especially important because such systems can incur substantial energy and cost overheads through iterative interactions, redundant tool use, and repeated model invocations.
Understanding this dimension is crucial for designing sustainable and cost-effective agentic frameworks, particularly as they transition from experimental prototypes to large-scale industrial deployment.

Our work addresses this overlooked aspect by providing the first end-to-end, hardware-level energy analysis of multi-turn agentic workflows powered by SLMs.
Unlike prior studies that isolate either model execution or code efficiency, we present a unified evaluation of both the computational cost and the functional effectiveness of SLM-based agentic systems originally designed for LLMs.

\section{Study Design}
\label{sec:study_design}

\begin{figure*}[htbp]
  \centering
  \includegraphics[width=\linewidth]{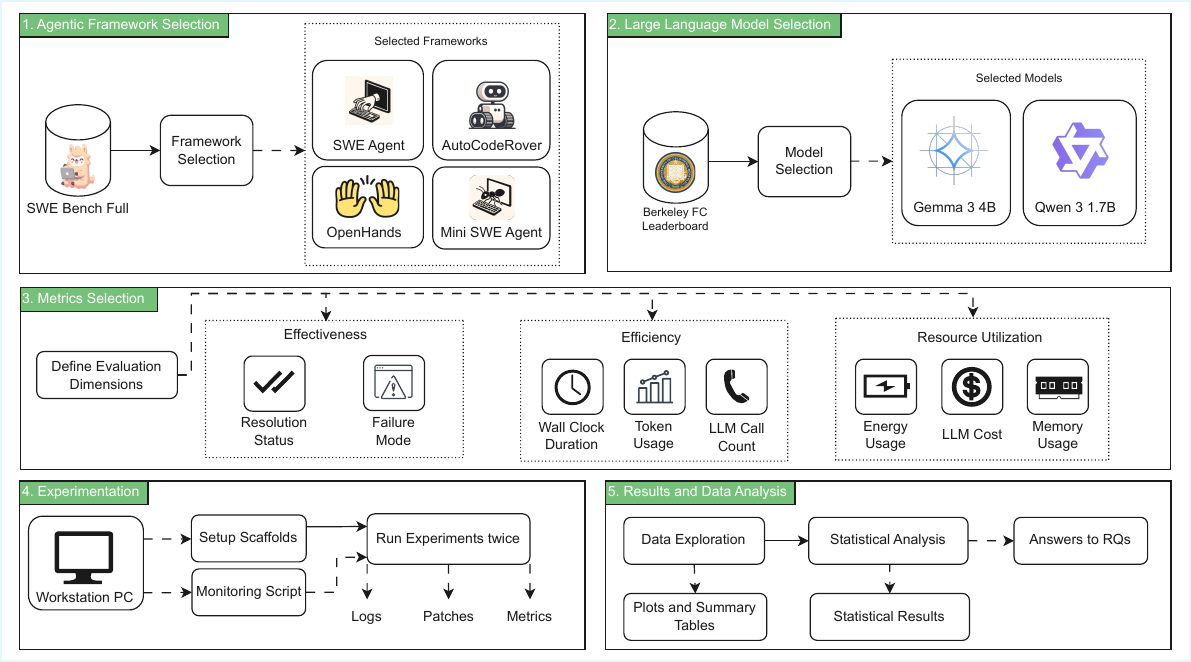}
  \caption{Overall study design showing the major phases.}
  \Description{A flow diagram illustrating the main phases of the study.}
  \label{fig:study_design}
\end{figure*}

Following the Goal Question Metric template by Basili et al. \cite{basili1994gqm}, the \textbf{goal} of this study is \textbf{to analyze} \textit{issue resolution frameworks powered by SLMs} for the \textbf{purpose of} \textit{evaluating their energy efficiency and effectiveness} from the \textbf{point of view of} \textit{developers} in the \textbf{context of} \textit{software development}. This goal is achieved by answering the following research questions.

\noindent\textbf{$RQ_1$ -- How do different agentic frameworks compare in terms of energy efficiency and task resolution effectiveness when paired with the same underlying SLM?}

\noindent This question aims to establish the relative performance of each framework, focusing on the trade-offs between successfully resolving an issue and the energy consumed to do so. By controlling for the model and hardware, we can isolate the impact of the framework's architecture on overall efficiency.

\noindent\textbf{$RQ_2$ -- To what extent do framework-level architectural choices account for observed differences in energy consumption and runtime when using SLMs?}

\noindent This question investigates the root causes behind any performance differences identified in $RQ_1$.

\noindent Figure \ref{fig:study_design} outlines the main phases of the study, which are detailed in the following sections. For independent verification, we provide a full replication package including the source code, logs, and measurements for each phase.\footnote{\url{https://github.com/sa4s-serc/swenergy}}

\subsection{Agentic Frameworks and Benchmark Selection}
We evaluate four leading agentic frameworks: \textit{SWE-Agent}, \textit{OpenHands}, and \textit{AutoCodeRover}. As of October 2025, the first three systems ranked among the top three performers on the SWE-bench Full leaderboard\footnote{https://www.swebench.com}, making them representative of state-of-the-art approaches to autonomous issue resolution. \textit{Mini SWE Agent} was additionally chosen from the SWE Bench Verified leaderboard for its simplicity. Each embodies a distinct architectural philosophy, allowing us to examine how different design choices affect energy consumption. The following paragraphs provide additional background on each of them.

\textit{SWE-Agent} introduces a specialized agent-computer interface (ACI) with curated commands for repository navigation, file inspection, and code editing. It employs ReAct-style prompting~\cite{yao2023react}, where each iteration consists of thought, action, and observation until the issue is resolved or a stopping condition is reached.

\textit{OpenHands} provides a general-purpose framework with a broader action space including shell commands, file operations, and multi-agent collaboration. It supports sandboxed Docker execution environments and accommodates diverse architectures from simple ReAct loops to complex planning approaches.

\textit{AutoCodeRover} decomposes issue resolution into three sequential phases: fault localization, context retrieval, and patch generation. This structured pipeline minimizes unnecessary LLM calls and reduces context accumulation by focusing on relevant code portions early in the workflow.

\textit{Mini SWE Agent} is a minimal variant of \textit{SWE-Agent} executing within a sandboxed Docker container with a single bash access point, relying solely on the model's internal reasoning to sequence commands, reducing orchestration overhead.

For our evaluation dataset, we use \textit{SWE-bench Verified Mini}, a curated 50-task subset of SWE-bench Verified. This benchmark preserves the distribution of difficulty, repository diversity, and test pass rates observed in the full 500-task dataset while reducing storage requirements from approximately 130 GB to 5 GB~\cite{hobbhahn2025swebenchverifiedmini}. The reduced scale enables intensive replication while maintaining representativeness.

\subsection{SLMs Selection}
To ensure representativeness while maintaining feasibility on consumer grade hardware, we restrict our study to open-weight SLMs with no more than four billion parameters. Model selection was guided by three criteria: (1) strong performance on public leaderboards within the $\leq$4B parameter range, (2) architectural diversity across model families, and (3) support for stable local inference.
We selected two models that jointly satisfy these conditions: \textit{Gemma-3 4B} \cite{gemma2024} and \textit{Qwen-3 1.7B Instruct} \cite{qwenfunc2025}. Both were among the top-ranking models fitting these criteria on the Berkeley Function-Calling leaderboard \cite{patil2025bfcl}.

\textit{Gemma-3 4B} is an instruction-tuned model with 4 billion parameters, optimized for general coding and reasoning. It provides a baseline for conventional instruction-tuned architectures on consumer-grade hardware. \textit{Qwen-3 1.7B Instruct}, with 1.7 billion parameters, is smaller and more memory-efficient, enabling faster inference and lower energy usage, while remaining competitive in reasoning and coding tasks. Comparing the two allows us to study whether framework efficiency can offset reduced model capacity in multi-turn workflows.

\subsection{Metrics Selection}
To provide a holistic view of agent performance, we established a comprehensive set of metrics through a structured selection process. Our approach began by defining three core dimensions for evaluation: \textbf{effectiveness}, \textbf{efficiency}, and \textbf{resource utilization}. We then cataloged all measurable outputs from a run, including status, time, energy, tokens, memory, and cost, ensuring each candidate metric was compatible with the SWE-bench benchmark and common evaluation practices. After validating our instrumentation sources (i.e., Intel's RAPL for CPU energy and NVIDIA's NVML for GPU energy) for accuracy, we finalized the metric set.

This process resulted in the following metrics, grouped by the evaluation dimensions they address:

\subsubsection{Effectiveness Metrics}
This group measures the agent's ability to successfully resolve the given software issue.

\begin{itemize}
    \item \textbf{Resolution Status:} A binary outcome (\textit{pass} or \textit{fail}) for each run, determined by the official SWE-bench evaluation script. A run passes only if the agent-generated patch correctly resolves the issue and all associated tests pass.
    
    \item \textbf{Failure Mode:} For failed runs, the reason for termination is categorized according to the Multi-Agent System Failure Taxonomy~\cite{cemri2025multi}.
\end{itemize}

\subsubsection{Efficiency Metrics}
This group quantifies the speed and computational effort of the agent's solution process.

\begin{itemize}
    \item \textbf{Wall-Clock Duration:} Total elapsed time (minutes) from initiation to completion or timeout. This measures end-to-end task latency.
    
    \item \textbf{Token Usage:} The sum of input (prompt) and output (completion) tokens processed by the LLM across all calls within a run. This serves as a proxy for the computational load on the model and correlates with API costs.
    
    \item \textbf{LLM Call Count:} The total number of model invocations during a run. This metric reflects the "chattiness" and complexity of the framework's orchestration strategy.
\end{itemize}

\subsubsection{Resource Utilization Metrics}
This group captures the hardware and financial costs incurred during a run.

\begin{itemize}
    \item \textbf{Total Energy Consumption:} The cumulative CPU+GPU energy per run, reported in kilojoules ($\mathrm{kJ}$), measured via RAPL (CPU) and NVML (GPU).
    
    \item \textbf{Peak Memory Usage:} The maximum Resident Set Size (RSS) for system RAM and the maximum allocated VRAM in gigabytes ($GB$) observed during a run. This indicates the minimum hardware memory requirements for deployment.
    
    \item \textbf{Cost:} The API costs for each LLM were multiplied by the tokens used in each run to estimate the inference cost, using per-token pricing from their public APIs (as of October 2025) for both Gemma\footnote{https://aistudio.google.com} and Qwen\footnote{https://www.alibabacloud.com/help/en/model-studio/models}.
\end{itemize}

\section{Experimental Procedure}

Our experimental procedure is designed to ensure control, reproducibility, and measurement accuracy.

\subsection{Controlled Environment}
All experiments were executed on a dedicated workstation with a fixed hardware configuration (Intel Xeon w3-2435 CPU, 32 GB RAM, NVIDIA RTX A2000 16 GB GPU). To prevent measurement interference, each run was run in complete isolation, with no other user or non-essential system workloads active. We recorded the machine's idle draw (51.69 W CPU, 2.70 W GPU) and subtracted this baseline from every RAPL and NVML trace to isolate agent-induced energy usage. Both models were loaded in FP16 for memory efficiency. Peak VRAM usage was 11 to 12 GB for Gemma-3 4B and 8 to 9 GB for Qwen-3 1.7B, within our 16 GB limit. All experiments used a 32K token context window to ensure fair comparison.

\subsection{Execution and Instrumentation}
In total, we performed 1,200 experimental runs. Each run involved executing one framework and model pair on a single issue from the \textit{SWE-bench Verified Mini dataset}. Runs were executed with a 30-minute wall-clock timeout and a 32K token context window limit. A custom monitoring script, launched concurrently with the agent, collected resource data throughout each run.

\subsection{Repetitions and Verification}
To account for variability and ensure the stability of our results, we performed three full repetitions for each framework and model configuration. This yielded 150 data points, with 3 repetitions for each of the 50 issues, for each experimental condition. The official SWE-bench evaluation harness was used to independently verify the generated patch and assign a final pass or fail status. All raw logs and computed metrics were stored in a structured JSON format to ensure transparency and facilitate downstream analysis.

\subsection{Qualitative Log Analysis}
\label{sec:qualitative_log}
To understand behavioral patterns underlying the quantitative results, we analyzed execution logs from the first run of each framework-model configuration (200 runs total).
We used Gemini 2.5 Flash to assist in processing the logs, identifying: (i) the sequence of actions leading to each failure modes (context loss, reasoning mismatch, etc., as per the MAST taxonomy \cite{cemri2025multi}), and (ii) specific behavioral patterns such as repetitive failed commands, verbose output generation, and inefficient information-gathering strategies.

\section{Results}

Our experiments reveal a significant trade-off between task-resolution effectiveness and energy efficiency, driven primarily by the architectural choices of each agentic framework. The quantitative findings, summarized in Table~\ref{tab:perf_metrics_minutes}, are organized by our first two research questions.

\subsection{RQ1: Comparing Framework Efficiency and Task Success}

Across all 150 runs, task resolution rates were exceptionally low. The only configuration to achieve any success was \textit{AutoCodeRover} paired with Qwen 1.7B, resolving an average of 2 out of 50 tasks, a 4\% success rate. However, this minimal success came at a steep price. \textit{AutoCodeRover} was, by a significant margin, the most resource-intensive framework, consuming a mean of $216.21\,\mathrm{kJ}$ (Gemma) and $208.42\,\mathrm{kJ}$ (Qwen) per run after idle subtraction.

\begin{figure}[htbp]
  \centering
  \includegraphics[width=\linewidth]{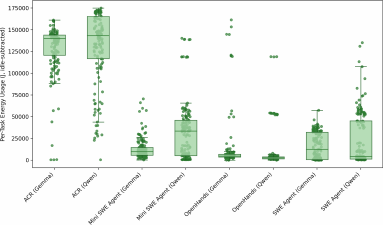}
  \caption{Energy usage distribution across different scaffolds and models.}
  \label{fig:energy_usage}
\end{figure}

In stark contrast, \textit{OpenHands} was the most energy-efficient framework, with its Gemma configuration consuming just $23.05\,\mathrm{kJ}$ on average (and $23.33\,\mathrm{kJ}$ for Qwen). Yet, this efficiency was coupled with a 0\% resolution rate. This difficult trade-off is evident across all configurations: nearly all frameworks achieved zero resolutions while consuming varying amounts of energy. \textit{AutoCodeRover} stands alone with non-zero success but at a significantly higher energy cost. As illustrated in Figure~\ref{fig:energy_usage}, \textit{AutoCodeRover} configurations show the widest spread and highest median energy while \textit{OpenHands} remains tightly clustered below 30 kJ. \textit{SWE-Agent} and \textit{AutoCodeRover} exhibit erratic energy patterns with large interquartile ranges, reflecting inconsistent termination behavior across runs.

\begin{tcolorbox}[colback=greenanswerbox, colframe=black, boxrule=1pt, sharp corners]
\textbf{Answer to RQ1:} When using SLMs, the frameworks demonstrate a severe trade-off between efficiency and effectiveness. The only successful configuration (\textit{AutoCodeRover}) was also the most energy-intensive (mean of $208$--$216\,\mathrm{kJ}$). The most efficient framework, \textit{OpenHands} (mean of $\approx 23\,\mathrm{kJ}$), resolved zero issues, forcing a choice between minimal success at a high cost or complete failure at a lower cost.
\end{tcolorbox}
\subsection{RQ2: Architectural Choices as the Primary Driver of Cost}

\begin{figure}[htbp]
  \centering
  \includegraphics[width=\linewidth]{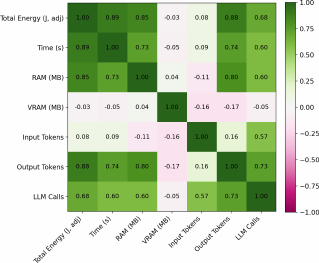}
  \caption{Correlation analysis across all frameworks.}
  \Description{Correlation matrix or plot showing relationships between various metrics across frameworks.}
  \label{fig:correlation}
\end{figure}

\begin{figure}[htbp]
  \centering
  \includegraphics[width=\linewidth]{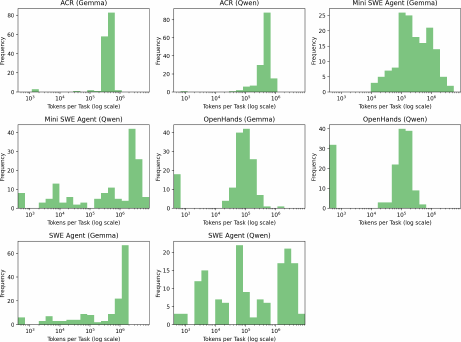}
  \caption{Token usage histograms for different frameworks.}
  \Description{Histograms showing the distribution of token usage across runs for each framework.}
  \label{fig:token_usage}
\end{figure}

The vast differences in resource consumption are directly linked to architectural patterns. The correlation heatmap (Figure~\ref{fig:correlation}) reveals Total Energy Usage is strongly correlated with Wall-Clock Duration ($R = 0.89$) and Output Tokens ($R = 0.88$). This confirms energy cost is driven by runtime and the ``chatty'' generation volume of ReAct-style agents (Figure~\ref{fig:token_usage}). In contrast, memory metrics show near-zero correlation, as memory is allocated statically upfront and does not vary with the computational workload.

\begin{figure}[htbp]
  \centering
  \includegraphics[width=\linewidth]{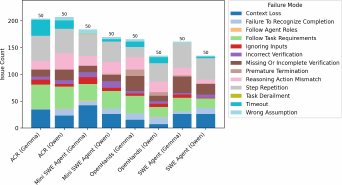}
  \caption{Failure reasons stacked bar chart.}
  \Description{Stacked bar chart depicting the distribution of failure modes for each framework.}
  \label{fig:failure_reasons}
\end{figure}

The failure analysis (Figure~\ref{fig:failure_reasons}) shows how architectures create distinct failure ``signatures'' from core SLM reasoning errors. \textit{AutoCodeRover} exhibits a balanced profile of failures like \texttt{Follow Task Requirements} and \texttt{Step Repetition}, where \texttt{Timeout} is a secondary \textit{consequence} of these unresolved loops, not the primary cause. In contrast, ReAct-style agents are dominated by \texttt{Step Repetition}, which directly causes frequent \texttt{Context Loss} failures as high token output exceeds the context window. Finally, \textit{OpenHands} shows a broad-spectrum failure signature, struggling across \texttt{Follow Task Requirements}, \texttt{Step Repetition}, and \texttt{Missing or Incomplete Verification}, suggesting its general-purpose design lacks specialized guardrails.

\begin{tcolorbox}[colback=greenanswerbox, colframe=black, boxrule=1pt, sharp corners]
\textbf{Answer to RQ2:} Energy consumption is primarily driven by runtime and output token usage ($R \approx 0.88$). Framework architecture dictates how SLM reasoning failures manifest: ``Chatty'' ReAct-style agents get stuck in repetitive loops, leading to high token counts and eventual \texttt{Context Loss}. Deliberative frameworks like \textit{AutoCodeRover} also suffer from reasoning loops, consuming high energy through long runtimes that often end in \texttt{Timeout} as a secondary consequence.
\end{tcolorbox}

\begin{table*}[tbh]
\centering
\caption{Performance Metrics Comparison Across All Repetitions. Energy columns report CPU/GPU draw after subtracting the measured idle baseline (51.69 W CPU, 2.70 W GPU).}
\label{tab:perf_metrics_minutes}
\footnotesize
\setlength{\tabcolsep}{3pt}
\begin{tabular}{
l
l
>\bfseries l
S[table-format=4.2]
S[table-format=4.2]
S[table-format=4.2]
S[table-format=5.2]
S[table-format=3.2]
S[table-format=5.2]
S[table-format=8.2]
S[table-format=2.2]
S[table-format=1.0]
}
\toprule
\textbf{Framework} &
\textbf{Model} &
\textbf{Statistic} &
{\textbf{Time (min)}} &
{\textbf{CPU (kJ)}} &
{\textbf{GPU (kJ)}} &
{\textbf{Total Energy (kJ)}} &
{\textbf{RAM (MB)}} &
{\textbf{VRAM (MB)}} &
{\textbf{Tokens}} &
{\textbf{Cost (USD)}} &
\multicolumn{1}{c}{\textbf{Resolved}}
\\
\midrule
\multirow{10}{*}{\textbf{AutoCodeRover}} &
\multirow{5}{*}{\textbf{Gemma}} & Min & 26.46 & 105.77 & 106.23 & 212.00 & 865.09 & 11623.81 & 392337.68 & 0.0077 & 0.0 \\
& & P50 & 27.20 & 107.49 & 107.75 & 217.99 & 882.45 & 11733.89 & 396991.10 & 0.0079 & 0.0 \\
& & Mean & 27.08 & 108.05 & 108.16 & 216.21 & 876.98 & 11724.22 & 405436.60 & 0.0081 & 0.0 \\
& & Max & 27.59 & 110.89 & 110.50 & 218.65 & 883.39 & 11814.96 & 426981.02 & 0.0085 & 0.0 \\
& & Std & 0.57 & 2.61 & 2.17 & 3.66 & 10.30 & 95.94 & 18802.53 & 0.0004 & 0.0 \\
\cmidrule(l){2-12}
& \multirow{5}{*}{\textbf{Qwen}} & Min & 22.91 & 111.25 & 91.01 & 202.26 & 793.82 & 8163.81 & 446642.64 & 0.1200 & 1.0 \\
& & P50 & 24.20 & 112.00 & 96.42 & 208.42 & 795.63 & 8170.28 & 463378.38 & 0.1262 & 2.0 \\
& & Mean & 23.79 & 113.72 & 94.70 & 208.42 & 795.45 & 8208.77 & 460457.63 & 0.1245 & \textbf{2.0} \\
& & Max & 24.26 & 117.91 & 96.67 & 214.58 & 796.89 & 8292.21 & 471351.88 & 0.1273 & 3.0 \\
& & Std & 0.76 & 3.65 & 3.20 & 6.16 & 1.55 & 72.33 & 12610.90 & 0.0039 & 1.0 \\
\midrule
\multirow{10}{*}{\textbf{Mini SWE Agent}} &
\multirow{5}{*}{\textbf{Gemma}} & Min & 3.12 & 12.00 & 8.97 & 20.73 & 99.07 & 11504.44 & 495699.42 & 0.0199 & 0.0 \\
& & P50 & 3.18 & 12.98 & 10.84 & 23.30 & 99.24 & 11505.47 & 555194.60 & 0.0223 & 0.0 \\
& & Mean & \textbf{3.20} & 13.54 & 10.23 & 23.41 & 99.21 & 11527.25 & 547710.07 & 0.0220 & 0.0 \\
& & Max & 3.30 & 15.63 & 10.86 & 26.19 & 99.32 & 11571.84 & 592236.18 & 0.0238 & 0.0 \\
& & Std & 0.09 & 1.88 & 1.09 & 2.73 & 0.13 & 38.62 & 48701.64 & 0.0020 & 0.0 \\
\cmidrule(l){2-12}
& \multirow{5}{*}{\textbf{Qwen}} & Min & 7.80 & 26.23 & 26.23 & 49.31 & 99.32 & 8285.59 & 1796615.00 & 0.2084 & 0.0 \\
& & P50 & 8.27 & 29.47 & 27.69 & 53.62 & 99.73 & 8286.01 & 1902548.94 & 0.2217 & 0.0 \\
& & Mean & 8.54 & 28.83 & 28.96 & 54.13 & 99.65 & 8285.93 & 1874298.96 & 0.2178 & 0.0 \\
& & Max & 9.54 & 30.80 & 32.95 & 59.44 & 99.89 & 8286.19 & 1923732.94 & 0.2232 & 0.0 \\
& & Std & 0.90 & 2.35 & 3.54 & 5.08 & 0.29 & 0.31 & 68104.98 & 0.0082 & 0.0 \\
\midrule
\multirow{10}{*}{\textbf{OpenHands}} &
\multirow{5}{*}{\textbf{Gemma}} & Min & 2.38 & 7.91 & 6.61 & 14.36 & 53.84 & 11030.14 & 92548.56 & 0.0038 & 0.0 \\
& & P50 & 4.11 & 14.16 & 7.17 & 20.43 & 81.22 & 11517.47 & 110753.42 & 0.0045 & 0.0 \\
& & Mean & 4.13 & \textbf{12.98} & 10.87 & \textbf{23.05} & \textbf{74.30} & 11427.27 & \textbf{106444.17} & \textbf{0.0043} & 0.0 \\
& & Max & 5.89 & 16.87 & 18.83 & 34.36 & 87.84 & 11734.19 & 116030.54 & 0.0047 & 0.0 \\
& & Std & 1.75 & 4.60 & 6.90 & 10.25 & 18.03 & 360.59 & 12319.82 & 0.0005 & 0.0 \\
\cmidrule(l){2-12}
& \multirow{5}{*}{\textbf{Qwen}} & Min & 6.37 & 17.09 & 5.37 & 20.56 & 69.50 & 8122.04 & 106271.78 & 0.0126 & 0.0 \\
& & P50 & 6.86 & 20.01 & 5.52 & 23.47 & 99.10 & 8163.81 & 111827.50 & 0.0133 & 0.0 \\
& & Mean & 6.89 & 20.18 & \textbf{5.59} & 23.33 & 89.40 & \textbf{8149.89} & 110228.31 & 0.0131 & 0.0 \\
& & Max & 7.44 & 23.44 & 5.86 & 25.95 & 99.60 & 8163.81 & 112585.66 & 0.0133 & 0.0 \\
& & Std & 0.54 & 3.18 & 0.25 & 2.70 & 17.24 & 24.12 & 3447.36 & 0.0004 & 0.0 \\
\midrule
\multirow{10}{*}{\textbf{SWE Agent}} &
\multirow{5}{*}{\textbf{Gemma}} & Min & 5.03 & 19.61 & 11.01 & 31.63 & 164.54 & 11504.44 & 707680.86 & 0.0277 & 0.0 \\
& & P50 & 5.47 & 20.61 & 12.99 & 32.60 & 181.66 & 11504.44 & 819349.52 & 0.0324 & 0.0 \\
& & Mean & 5.59 & 22.05 & 12.73 & 34.78 & 177.03 & 11711.76 & 788841.29 & 0.0311 & 0.0 \\
& & Max & 6.26 & 25.93 & 14.20 & 40.12 & 184.88 & 12126.42 & 839493.48 & 0.0333 & 0.0 \\
& & Std & 0.62 & 3.40 & 1.61 & 4.65 & 10.93 & 359.10 & 71004.97 & 0.0030 & 0.0 \\
\cmidrule(l){2-12}
& \multirow{5}{*}{\textbf{Qwen}} & Min & 8.85 & 27.17 & 18.44 & 41.80 & 210.66 & 8163.81 & 1177137.68 & 0.1349 & 0.0 \\
& & P50 & 8.94 & 30.25 & 19.31 & 45.32 & 216.11 & 8163.81 & 1238045.34 & 0.1413 & 0.0 \\
& & Mean & 9.02 & 29.84 & 19.22 & 44.87 & 215.27 & 8204.60 & 1250213.61 & 0.1428 & 0.0 \\
& & Max & 9.26 & 32.09 & 19.90 & 47.49 & 219.04 & 8286.19 & 1335457.80 & 0.1521 & 0.0 \\
& & Std & 0.21 & 2.49 & 0.73 & 2.87 & 4.25 & 70.65 & 79858.41 & 0.0087 & 0.0 \\
\midrule
\bottomrule
\end{tabular}
\end{table*}

\section{Discussion}
\label{sec:discussion}

Our quantitative results establish that under realistic hardware constraints, current agentic frameworks struggle to resolve software issues reliably or efficiently. To explain these outcomes, we conducted a qualitative analysis of the execution logs from the first run of each framework and model pair. This analysis, performed with the assistance of the Gemini 2.5 Flash model, revealed that the architectural flaws do not lie in the frameworks themselves, but in their implicit assumption of a highly capable reasoning engine. This core mismatch leads to several key lessons and has direct implications for both future research and current practice.

\subsection{Lessons Learned}

Our qualitative analysis of failure modes revealed critical lessons about the interaction between framework design and SLM limitations.

\textit{Lesson 1: Frameworks lacking SLM-aware safeguards waste energy on unproductive loops.} High resource consumption was a direct result of frameworks allowing the SLM to enter unproductive reasoning loops without intervention. \textit{AutoCodeRover}, the only framework with any success, frequently became trapped. Its rigid, multi-phase design proved brittle when the SLM failed to provide a correct input for the next stage. Lacking robust reasoning, the SLM would repeatedly submit the same failing test script or malformed API call. The framework, unable to recognize this failure pattern, allowed the SLM to burn time and energy in a futile loop. Similarly, \textit{OpenHands} agents often entered loops of inefficient exploration, issuing broad \texttt{grep} commands that returned massive, unfiltered outputs and overwhelmed their context windows.

\textit{Lesson 2: Low energy consumption is not synonymous with efficiency.} Conversely, low energy consumption did not necessarily indicate efficiency. In the case of \textit{Mini-SWE Agent}, a low energy profile was the result of two distinct failure patterns. First, a significant number of runs terminated prematurely when the agent's context management generated prompts that exceeded the SLM's token limit. The agent failed before it could begin substantive work. Second, our analysis uncovered a pattern of ``false positives,'' where the framework reported a task as ``successful'' despite producing a destructive or syntactically invalid patch (e.g., replacing a 2,000-line file with a single invalid line). This demonstrates that low energy metrics can conceal critical functional failures.

\textit{Lesson 3: Current agentic architectures are fundamentally mismatched with SLM capabilities.} The common thread is that current frameworks are designed as passive orchestrators that \emph{assume} a competent reasoning engine. They provide tools and wait for the model to use them correctly. This paradigm collapses when the model's reasoning is weak. The frameworks lack the necessary mechanisms to guide, scaffold, or correct a struggling SLM, leading directly to the high-consumption loops and low-consumption failures we observed.

\subsection{Implications for Research}

These lessons point to a clear need for a new class of agentic frameworks, aligning with the broader research agenda for environmentally sustainable AI practices~\cite{10.1145/3743095.3743099}. Future work must move beyond architectures that encourage brute-force interaction and toward those that foster intelligent, adaptive reasoning. Frameworks should integrate adaptive strategy management to detect and break unproductive reasoning loops, forcing the agent to revise its approach when progress stalls. They should also include guided exploration and context filtering mechanisms to prevent context overload, actively curating relevant tool outputs before passing them to the model. Finally, robustness demands the inclusion of independent verification layers, separating patch generation from evaluation. This would allow the framework to detect “false positives” and provide higher-fidelity feedback throughout the resolution process.

\subsection{Implications for Practice}

For developers and practitioners seeking to leverage SLMs for autonomous tasks, the findings offer practical guidance and caution. Plug-and-play substitution of an LLM with an SLM is ineffective and often counterproductive, as existing frameworks are not built for the limited reasoning capacity of smaller models. Practitioners must instead adopt or design frameworks explicitly tailored for SLMs. Moreover, evaluation should consider the total cost of ownership (TCO), not just the absence of API costs. Local execution may appear cheaper but often incurs heavy time and energy costs, such as the 24--27 minute mean runtimes for \textit{AutoCodeRover} configurations. Practitioners can use decision maps \cite{lago2019architecture}, to explicitly model these trade-offs and evaluate whether their own agentic architectures are sustainable across technical, economic, and environmental dimensions.Finally, given the frequency of “false positive” task completions, it is critical that all SLM-driven systems employ rigorous external verification, such as CI-based build and test validation, before any generated patches are accepted or presented for human review.

\section{Threats to Validity}
\label{sec:threats}

We identify several potential threats to the validity of our findings, categorized according to established guidelines for empirical studies.

\noindent\textbf{Internal Validity.} We mitigated this threat by executing all experiments on a single, dedicated machine in an isolated environment, with no other significant user or system processes running. However, minor, non-deterministic background OS activity could still have a marginal effect on measurements.

\noindent\textbf{Construct Validity.} Our measurements focus on primary consumers (CPU and GPU) via standard interfaces (RAPL and NVML). While this is a common and accepted practice, it does not capture the total "wall plug" energy, as consumption from RAM, storage, and other motherboard components is excluded. Furthermore, our effectiveness metric is the binary \textit{pass/fail} status from the SWE-bench harness, which, while standard, does not account for partial progress or the quality of a generated solution. For our qualitative analysis (Section \ref{sec:qualitative_log}), we used Gemini 2.5 Flash to process execution logs. This introduces a potential threat, as the AI assistant may influence pattern identification and interpretation of failure behaviors. To mitigate this, two of the authors randomly sampled and manually analysed the logs to verify the accuracy of the AI-assisted findings. We also provide complete raw logs in our replication package to enable independent re-analysis.

\noindent\textbf{External Validity.} Three main threats affect generalizability: First, we used the \textit{SWE-bench Verified Mini} benchmark. While representative, its 50 tasks are a small subset of real-world software engineering problems, and our findings may not generalize to different types or scales of issues. Second, our experiments were limited to two small, open-weight models. The observed performance and efficiency trade-offs could differ substantially with larger, more capable models (e.g., GPT-5 or Llama-3 70B), which are often used in state-of-the-art systems. Our choice was deliberate to focus on resource-constrained, local deployment scenarios. Third, all measurements were performed on a single hardware configuration. The energy profiles may vary on different CPU/GPU architectures (e.g., AMD or Apple Silicon).

\noindent\textbf{Conclusion Validity.} This relates to the statistical reliability of our conclusions. A significant threat arises from the extremely low task resolution rate across all configurations. With only one framework achieving any success, our conclusions about the resource consumption of failed runs are far more robust than those about successful ones. The correlations and comparisons are based on a dataset heavily skewed toward failure, limiting our ability to draw firm conclusions about the energy cost of a successfully resolved issue.

\section{Conclusion}
In this study, we investigated the energy efficiency and effectiveness of autonomous software engineering agents when powered by Small Language Models (SLMs). We conducted a controlled experiment evaluating four prominent agentic frameworks with two SLMs on the \textit{SWE-bench Verified Mini} benchmark, collecting detailed measurements of energy consumption and other resource metrics on a fixed hardware platform.

Our results highlight a significant gap: current agentic frameworks, designed for the strengths of large models, are ill-equipped to handle the reasoning limitations of SLMs. As expected, task resolution rates were extremely low. We found that while framework architecture is a primary driver of energy consumption, the SLM's capacity was the bottleneck for success. High-energy failures were often caused by frameworks allowing the SLM to enter unproductive loops, while low-energy consumption sometimes masked premature terminations or flawed outputs from the model.

This work reframes the challenge from simply making agents more accurate to designing them to be effective within realistic resource constraints. Future work must focus on developing frameworks that actively guide and assist SLMs. This includes creating mechanisms for adaptive strategy management, integrating tools for guided exploration, and establishing robust, independent verification protocols as standard. By treating the model's limitations as a core design constraint, the research community can build the next generation of autonomous agents that are not only effective but also sustainable and accessible for real-world, local deployment.

\begin{acks}
This research was funded by the ANRF Prime Minister Early Career Research Grant (ANRF/ECRG/2024/003379/ENS).
\end{acks}

\bibliographystyle{ACM-Reference-Format}
\bibliography{references}
\end{document}